\documentclass[twoside]{amsart}
\usepackage{latexsym}
\usepackage{amssymb,amsmath,amsopn}
\usepackage[dvips]{graphicx}   
\usepackage{color,epsfig}      

               {\begin{list}{}{\leftmargin#1\rightmargin#2}\item{}}%
               {\end{list}}
 
\newtheorem{thm}{Theorem}

\theoremstyle{definition}

\newtheorem{conj}{Conjecture}

\def\bea{\begin{eqnarray}}
\def\eea{\end{eqnarray}}

\begin{document}
\title[Soft cells, seashells]{Soft cells and the geometry of seashells}
\author[G. Domokos, A. Goriely,  \'A. G. Horv\'ath \and K. Reg\H os]{G. Domokos, A. Goriely,  \'A. G. Horv\'ath \and K. Reg\H os}

\address{G\'abor Domokos, HUN-REN-BME Morphodynamics Research Group and Dept. of Morphology and Geometric Modeling, Budapest University of Technology,
M\H uegyetem rakpart 1-3., Budapest, Hungary, 1111}
\email{domokos@iit.bme.hu}
\address{Alain Goriely, Mathematical Institute, University of Oxford}
\email{goriely@maths.ox.ac.uk}
\address{\'Akos G. Horv\'ath, HUN-REN-BME Morphodynamics Research Group and Department of Algebra and Geometry, Budapest University of Technology and Economics, H-1111 Budapest, M\H uegyetem rkp 3.}
\email{ghorvath@math.bme.hu}
\address{Krisztina Reg\H os, HUN-REN-BME Morphodynamics Research Group and Dept. of Morphology and Geometric Modeling, Budapest University of Technology,
M\H uegyetem rakpart 1-3., Budapest, Hungary, 1111}
\email{regoskriszti@gmail.com}

\begin{abstract}
A central problem of geometry is the tiling of space with simple structures. The classical solutions, such as triangles, squares, and hexagons in the plane and cubes and other polyhedra in three-dimensional space are built with sharp corners and flat faces. However, many tilings in Nature are characterized by shapes with curved edges, non-flat faces, and few, if any, sharp corners.  An important question is then to relate  prototypical sharp tilings  to  softer natural shapes. Here, we solve this problem by
 introducing a new class of shapes, the \textit{soft cells}, minimizing the number of sharp corners and filling space as \emph{soft tilings}. We prove that an infinite class of polyhedral tilings can be smoothly deformed into soft tilings and 
 we construct the soft versions of all Dirichlet-Voronoi cells associated with point lattices in two and three dimensions. Remarkably, these ideal soft shapes, born out of geometry, are found abundantly in nature, from cells to shells.
\end{abstract}
\maketitle

\pagebreak
\tableofcontents
\pagebreak
\section{Introduction}

The quest to find \emph{tilings}, i.e. space-filling patterns consisting of non-overlapping, finite domains, started  more than ten thousand years ago with the advent of masonry walls. However, tilings are much older than that: they are an integral part of Nature. Here, we describe a new class of space-filling patterns  called \emph{soft tilings} with highly curved cells which minimize the number of sharp corners. To motivate this concept, we first briefly review simpler tilings.
\subsection{From Plato to Plateau: cells with flat and slightly curved faces}
  The first geometric theory of tilings dates back to Plato \cite{Senechal_tetrahedra} who claimed that the five regular polyhedra, the \emph{ Platonic solids}, fill space without gaps, forming the four fundamental substances: earth, air, fire and water, while the fifth solid (the dodecahedron) is the building block of the cosmos.  Plato's views were soon found to be flawed. Indeed, Aristotle claimed that only the cube and the tetrahedron can fill space without gaps. The latter statement is incorrect for the regular tetrahedron. However, there exist space-filling tetrahedral honeycombs \cite{Senechal_tetrahedra} to which we will return later. Plato's idea was resurrected in the study of non-Euclidean honeycombs \cite{coxeter1956} where all Platonic solids fill space. 

Plato's idea and Aristotle's refinement lead, ultimately, to the fundamental concept of the solid angle which proved to be an essential tool in describing the combinatorial properties of  \emph{convex tilings} \cite{Senechal, schneider2008stochastic} which are also \emph{polyhedral tilings}, filling space by convex polyhedra, without gaps and overlaps.

\begin{figure}[h!]
\begin{center}
\includegraphics[width=\columnwidth]{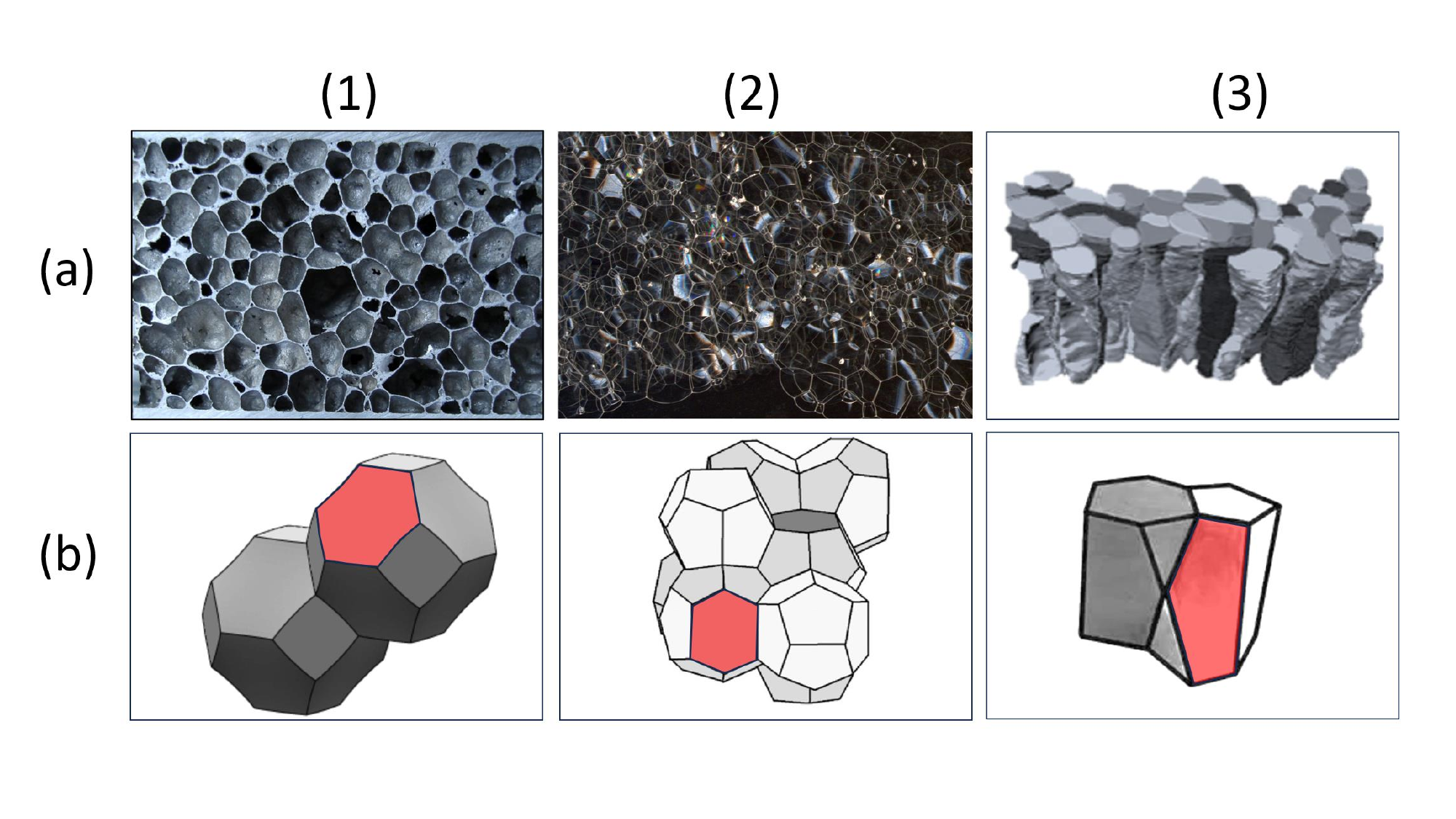}
\caption{Examples for \emph{slightly curved} polyhedric tilings. Upper row: natural examples. Lower row: geometric models. One slightly curved face highlighted on each tiling.  (a1) Metal foam (source: Wikimedia Commons) (a2) Liquid foam (source: Wikimedia Commons).(a3) ephitalia tissue \cite{scutoid_ephitalia_image}. (b1) The Kelvin structure: a monohedric tiling. (b2)  The Weaire-Phelan structure: a polyhedric tiling with 2 cells. Source \cite{Bitsche_2005, daxner_2006}. (b3) Tiling with scutoids \cite{scutoid_gomez}.}\label{fig:1}
\end{center}
\end{figure}

 Polyhedral tilings provide good models for phenomena ranging from fragmentation processes \cite{Plato}  and the emergence of aeolian ridges on Mars \cite{Nagle2021} to supramolecular  patterns in monolayers \cite{Regos2023}. However, in many cases the geometry of the natural tessellation appears to be more complex than a polyhedral tiling. Yet, these more intricate structures can still often be  related to polyhedral tilings: two tilings are called \emph{combinatorially equivalent} if there is a bijection between their vertices, edges, and faces which preserves inclusion and adjacency. For instance, the shape of foams, controlled by Plateau's Laws \cite{Ball}, can be modeled by tilings which are combinatorially equivalent to a polyhedral tilings. However, in this case, faces may not be planar and edges may not be straight. These properties  are well reflected in the geometric foam models: the Kelvin structure  \cite{Weaire2009} and its improved version, the Weaire-Phelan structure \cite{Weaire1994}, shown in Figure \ref{fig:1}. The small deviation in curvature observed in foams is the result of physical constraints, not the space-filling constraint. This idea was taken one step further in the description of the closely packed cell system of ephitelia tissue \cite{scutoid_ephitalia_image}. This geometric model, shown in Figure \ref{fig:1} (3b), also has cells with curved faces and edges. While these cells are still \emph{individually} combinatorially equivalent to polyhedra, the tiling as a whole is not combinatorially equivalent to any polyhedral tiling, showing that in this case curvature is a result of the constraint to fill space.
Despite having different reasons for being curved,  all three mentioned structures only differ \emph{slightly} from polyhedral tilings in the sense that edges and faces meet transversely (angles both between faces and between edges are nonzero) and curvature radii along edges are very large compared to the size of the cell. This small deviation from polyhedral tilings may be the reason why such tilings are accepted as suitable models. Strikingly, in the theory of tilings no new category has been created to capture phenomena where the curvature of cells and edges my be large enough to play a central role.

\subsection{Large curvatures and the intuitive concept of soft tilings}
\begin{figure}[h!]
\begin{center}
\includegraphics[width=\columnwidth]{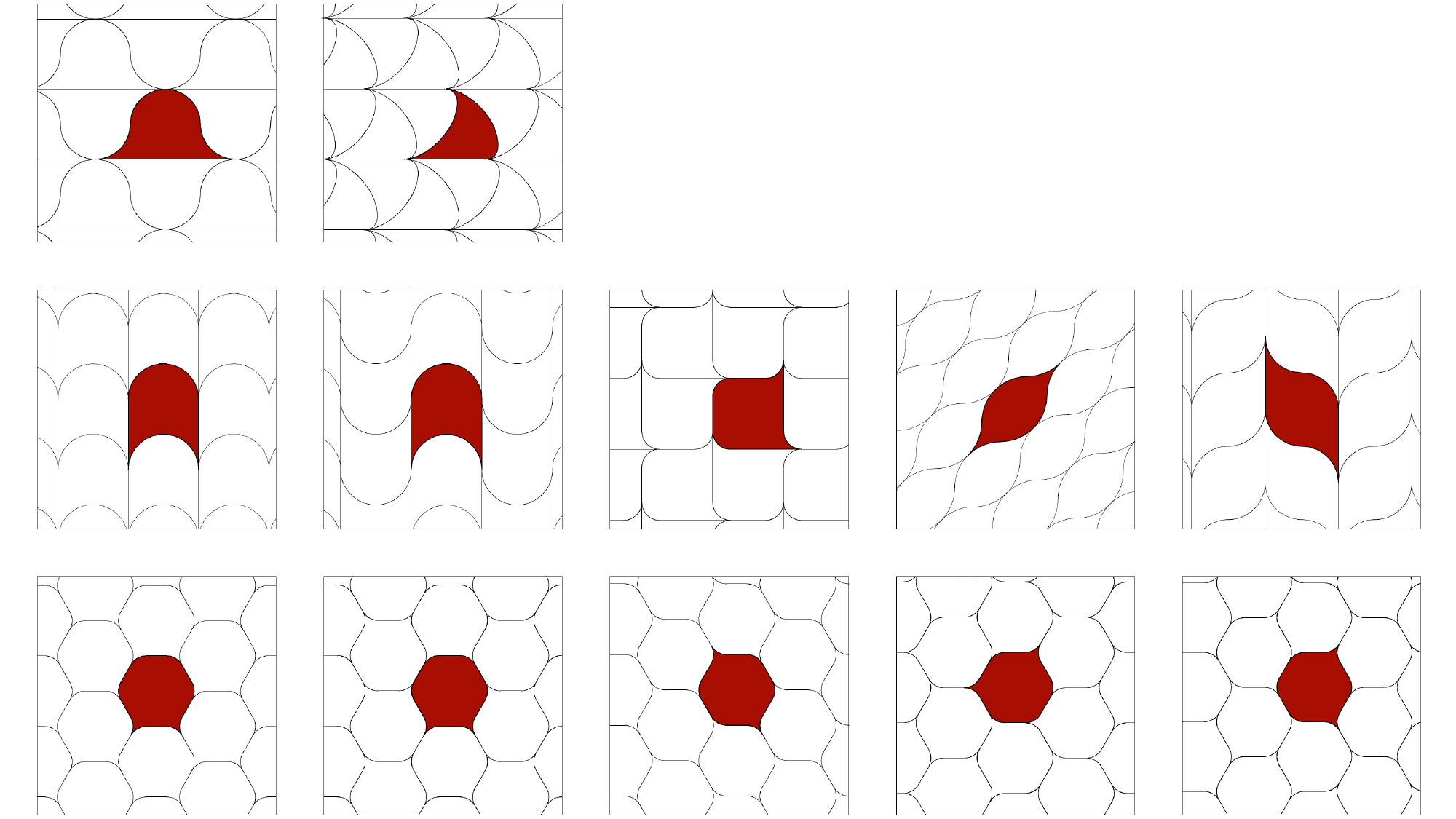}
\caption{Soft tilings in the plane. We show soft monohedric tilings which are combinatorially equivalent to monohedral tilings with regular polygons. Each row shows combinatorially equivalent soft tilings, corresponding to regular triangulation (first row), the rectangular grid (second row) and the hexagonal honeycomb (third row). If, beyond combinatorial equivalence classes, we also distinguish between sharp and soft corners then we arrive at the 12 tilings shown in the figure. In each tiling one soft cell is highlighted in red color.   We remark that the last two rows show soft tilings which are combinatorially equivalent to  Dirichlet-Voronoi mosaics on point lattices \cite{ghorvath_dirichlet}.} \label{fig:2}
\end{center}
\end{figure}

If we keep the combinatorial structure  of  faces and vertices but we allow tangencies (zero angles) between them and we also allow faces and edges
where curvature radii are comparable to the cell size, we enter an entirely new domain where radically new geometric features emerge. In particular, space-filling cells with tangencies and/or large curvatures may have \emph{fewer corners} than polyhedral simplices and such cells (which, depending on the number of corners we will call either \emph{softened} or \emph{soft}) do emerge in Nature. Figure \ref{fig:2} illustrates 2D examples of cells with curved boundaries which have only two corners and 3D examples will be discussed in the next section.

To create a geometric framework for  this generalization of polyhedral tilings  we introduce the concept of \emph{polyhedric tilings} which includes, beyond polyhedral tilings, tilings with cells having zero internal angles, cells having strongly curved faces and/or edges (see SI, section 1A for a formal definition).  
We do not consider here pathological tilings by assuming for the rest of the paper that both  polyhedral and polyhedric tilings  are \emph{normal} and \emph{balanced}, i.e. the cell diameter  has uniform upper and lower bounds and the averages of cell and nodal degrees exist \cite{schneider2008stochastic}.


If a tiling consists of identical cells, we refer to the cells and to the tiling as \emph{monomorphic}. In particular, if those cells are polyhedra then the tiling is \emph{monohedral} and its curved generalization is called \emph{monohedric}. There are infinitely many monohedral and monohedric tilings. Examples of monohedral tilings include the cubic grid and tilings with tetrahedral cells \cite{Senechal_tetrahedra}. An example of a monohedric tiling is the Kelvin structure of  Figure \ref{fig:1} (1b).

While the combinatorial properties of tilings have been investigated in detail, less attention was paid to the smoothness of the cells.  In  two dimensions, shapes with at least $C^1$-smoothness do not fill space \cite{twovertex}.
Since the planar sections  of 3D tilings are 2D tilings and the generic planar sections of $C^1$-smooth 3D objects are $C^1$-smooth 2D objects, this property of 2D tilings also implies that a 3D tiling can not consist entirely of $C^1$-smooth tiles. 

\begin{figure}[h!]
\begin{center}
\includegraphics[width=\columnwidth]{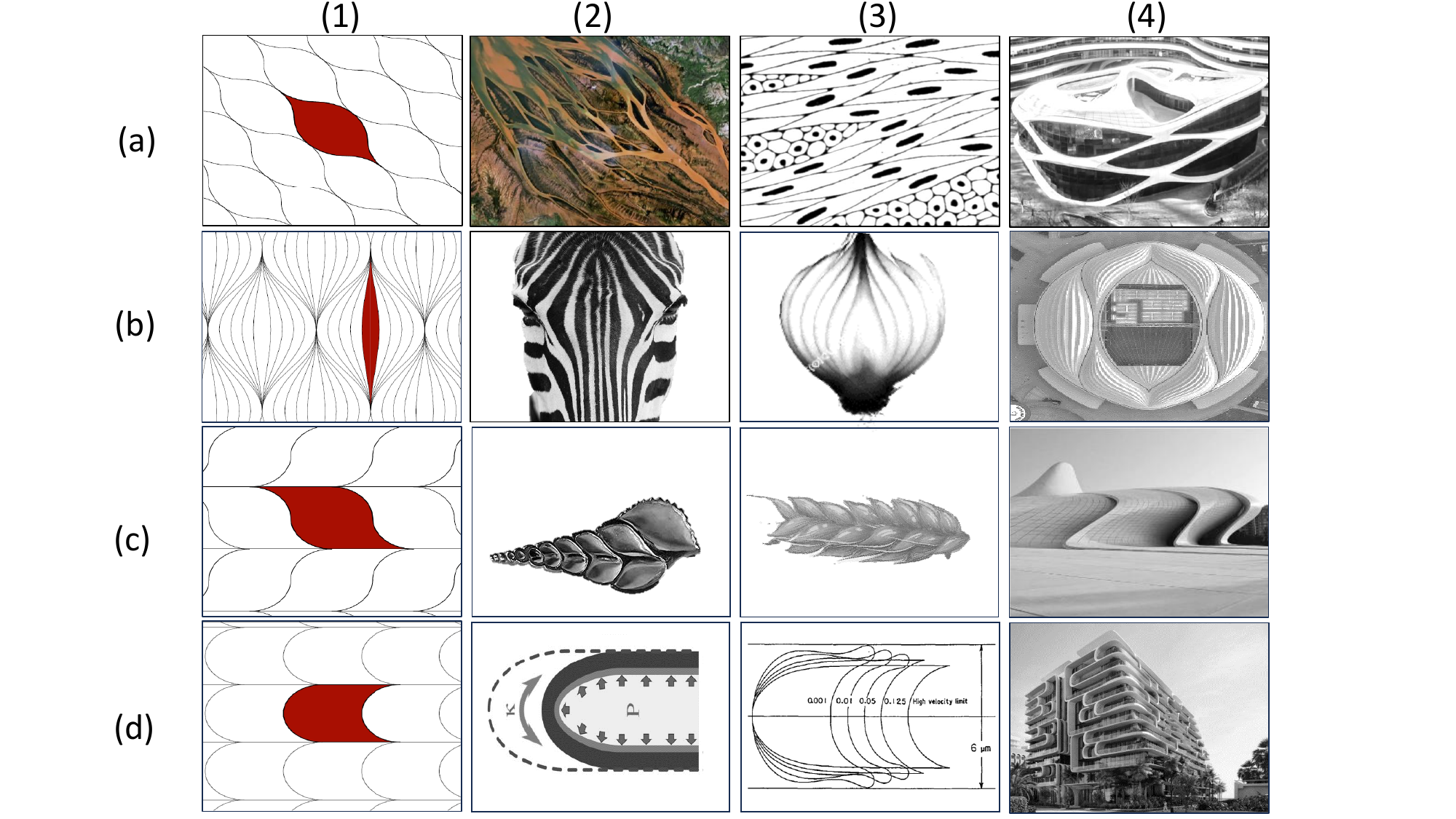}
\caption{Soft tilings in the plane: examples from nature and architecture. Column (1):  examples of monohedric, soft tilings with $v^{*}=2$ cells.(Remark that (a1),(c1) and (d1) are monohedric normal tilings, combinatorially equivalent to tilings with regular polygons, shown in Figure \ref{fig:2}, while (b1) is not a normal tiling.)  Columns (2) and (3): examples in Nature where these patterns emerge. (a2)  Betsiboka River estuary in northwestern Madagascar. (b2) Zebra stripes. (c2) Cross section of see shell. (d2) Geometric model of tip growth in algae \cite{algae_tipgrowth} (a3) Smooth muscle tissue. (b3) Cross section of onion. (c3) Wheat awn (d3) meridian section of blood cell \cite{secomb1991red}. Column (4):works of architect Zaha Hadid. (a4) Galaxy Soho, Beijing. (b4) Football stadium, Quatar. (c4) Heydar Aliev Center, Baku (d4) Design for condominium in Surfside, Florida (2023).}\label{fig:3}
\end{center}
\end{figure}

Since piecewise smooth shapes, such as squares  and cubes, do fill space, and smooth shapes do not, a natural question is \emph{how smooth} space-filling shapes can be. The answer to this question  depends on how we measure smoothness. Using the concept of polyhedric tilings we can assign a smoothness-related measure  to monohedric cells and try to minimize this measure among possible space-filling shapes.  Since a monohedric cell cannot be smooth,  there exists a set of non-smooth points on its boundary. Just as the smoothness of a point has various degrees (by counting the number of existing derivatives), we can define levels of \emph{non-smoothness} by counting the codimension of smooth manifolds containing the point. The basic idea is  to reduce the number of points with highest level of non-smoothness.

To be more precise, let us call a boundary point $p$ of a cell a \emph{corner} of the cell  if  no smooth curve on the boundary contains $p$. We denote the number of corners of a cell by $v^{\star}$ and  we define $v^{\star}_{min}$ as the minimal number of corners that a cell of a monohedric tiling may have \cite{twovertex}. Both the $d=2$ dimensional Euclidean plane and the $d=3$ dimensional Euclidean space have monohedral simplicial (triangular and tetrahedral, respectively) tilings \cite{goldberg} with $v^{\star}=d+1$. Therefore, for $d=2,3$ we have $v^{\star}_{min} \leq d+1$ and we will show that, in fact, this inequality is strict. The quantity $v^{\star}_{min}$ was defined using the concept of monohedric tilings, however, we can also apply  the very same quantity to classify general polyhedric tilings and cells.  In $d$ dimensions, we will call cells with $v^{\star} \leq v^{\star}_{min}$ corners \emph{soft cells} and cells with $v^{\star}_{min} < v^{\star}<d+1$ corners \emph{softened cells.}  Analogously, we call a tiling consisting entirely of softened (soft) cells  a \emph{softened tiling (soft tiling)}, for more detail, see SI, section 1A. 


\section{Soft tilings and soft cells}

\subsection{Soft tilings in two dimensions}
In two dimensions, corners are equivalent to the vertices along the boundary and it is known   that on the Euclidean plane 
$v^{\star}_{min}=2$ \cite{twovertex}.Hence, in 2D  we do not have softened cells and softened tilings. Figure \ref{fig:2} shows the monohedric
soft tilings which are combinatorially equivalent to monohedral tilings with regular polygons
which also includes all Dirichlet-Voronoi tilings on point lattices in the plane \cite{ghorvath_dirichlet}.  Minimizing the number of corners may be seen as a guiding principle along which natural tilings evolved. Indeed, soft planar tilings emerge not only in natural patterns such as smooth muscle cells, seashells compartment and zebra stripes, they have been repeatedly used in the works of visionary architect Zaha Hadid (a.k.a. the 'Queen of Curves'), who abandoned the rectangular grid and created futuristic designs with curved facades. Figure \ref{fig:3} shows some soft tilings in two dimensions, how they appear in nature and in the work of Hadid.

\subsection{Soft tilings in three dimensions }

Unlike the two dimensional case, in 3D Euclidean
space one can build a monohedric cell without any corners \cite{twovertex}. Therefore,  the absolute minimum $v^{\star}_{min}=0$ can be reached which leaves
the range $0 < v^{\star} < 4$  for softened (but not soft) cells and tilings unexplored. We now proceed to study both soft and softened tilings.

As a stepping stone towards general soft 3D cells, and also serving as a link to applications in the geometry of seashells, we first we define  the concept of $z$-cells. A prism is generated by a simple, closed, piecewise smooth curve $b_0$ in the plane $P$ and the family of  all lines orthogonal to $P$ and intersecting $b_0$. We will refer to $b_0$ as the \emph{base} of the prism and we will call planar sections orthogonal to $P$ \emph{meridian sections}.  We call a compact shape a \emph{$z$-cell} if its copies fill a prism without gaps and without overlaps (see Materials and Methods, subsection \ref{ss:zcells} and SI, sec.3A.)  We can distinguish between 4 types of $z$-cells, based on two binary properties:
if  the base of the prism is a monohedric tile in the plane  then the $z$-cell is space-filling
(otherwise it is not space-filling). If the cell has no sharp vertices then it is soft (otherwise it is not soft). To the previous two, we add a third, independent binary category, applicable to any compact shape, determining whether or not it is a $z$-cell. These three  binary properties generate the 8 categories shown in Figure \ref{fig:4} (see Materials and Methods, subsection \ref{ss:softness} and SI, sections 2B, 3 and 4A).

\subsection{The edge bending algorithm and the main result}
Monohedric soft cells can be constructed from monohedral cells by the \emph{edge bending algorithm}: We start with a polyhedral tiling and smoothly bend each edge  in such a manner that at each node all half-tangents are aligned. In this process \emph{corners are converted into points of edges}, i.e. into points at which all smooth  curves of the boundary have identical tangents.  

For example, consider the cubic grid, a space-filling polyhedral tiling consisting of cubes as cells, see Figure \ref{fig:4}(c3) and Figure \ref{fig:7}(1b).  Three edges meet at any corner, their half-tangents pairwise at right angles (Figure \ref{fig:7}(1c)). As we apply the edge bending algorithm, we keep
the vertical half-tangent frozen (see Figure (Figure \ref{fig:7}(1)) and bend the other two edges so their half-tangents become also vertical. In this manner all three half-tangents are aligned and the corner disappears (Figure \ref{fig:7}(1d)). Intuitively, one can achieve softened and soft tessellations by the edge bending algorithm, the details of which we discuss in Materials and Methods,
subsection \ref{ss:edgebending} as well as in the SI, Sec.1.

The starting object for the edge bending algorithm  is a convex (polyhedral) tiling and we will call the execution of the algorithm \emph{successful} if it results in a soft tiling. Whether or not the edge bending process is successful for any arbitrary initial tiling is unknown. Edge bending is a highly complex geometric process transforming  edges and faces while developing possibly large curvatures. Surprisingly, we show that a purely combinatorial condition  on the initial tiling is sufficient for the algorithm to be successful. This condition hinges on three classic concepts which we briefly explain. Every polyhedron  $P$ is associated with a \emph{dual polyhedron $P'$}  in such a manner that  the vertices of $P$ correspond to the faces of $P'$ and vice versa and the edges between pairs of vertices of $P$ correspond to the edges between pairs of faces of $P'$ and vice versa. For example, the dual of the cube is the octahedron and the tetrahedron is a self-dual polyhedron. If $P$ is convex, so is $P'$. The second classic concept is a polyhedron's \emph{Hamiltonian circuit}:  a cyclic path along the edges of a polyhedron that visits each vertex exactly once. The condition for the non-existence of a Hamiltonian circuit for a given polyhedron depends on the number faces, edges and vertices. Necessary conditions (lower bounds) are known \cite{barnette_Hamiltonian_circuit}, but the exact minima for the number of faces, edges and vertices so that the polyhedron does not have a Hamiltonian circuit are not known, not even for simple polyhedra where three edges meet at every vertex. 
The third important notion is the concept of a \emph{vertex polyhedron}.
The tiles of polyhedral tilings are polyhedra. The nodes of the tiling are points where the vertices of polyhedral cells overlap and the degree $n_o$ of the node $o$ is equal to the number of polyhedra the vertices of which overlap at $o$. If we consider the intersection of the tiling with a small sphere centered at $o$ then we obtain a spherical polyhedron $P_o$ with $n_o$ faces. We call the Euclidean polyhedron (with straight edges and flat faces) which is combinatorially equivalent to $P_o$ the \emph{vertex polyhedron} at the node $o$. For example, in a regular cubic grid we have $n_o=8$, vertex polyhedra are octahedra because 8 cubes meet at every node. 
Using these three key concepts, we give the necessary condition for the edge bending algorithm to be successful:

\begin{thm}\label{thm:main}
Let $M$ be a balanced, normal convex tiling and let $\mathcal{V}(M)$ denote the set of the duals of vertex polyhedra in $M$. If  every polyhedron $P \in  \mathcal{V}(M)$ has a Hamiltonian circuit, then there exists a soft polyhedric tiling $M'$ which is combinatorially equivalent to $M$. 
\end{thm}

As an example, consider the cubic grid with the octahedron as its vertex polyhedron. The dual of this vertex polyhedron is the cube. Since there exists a Hamiltonian circuit on the cube (inset of Figure \ref{fig:4}(c3)), Theorem \ref{thm:main} guarantees that there exists a a soft polyhedric tiling which is combinatorially equivalent to the cubic grid. The cell of one particular example, which happens to be monohedric, is shown in Figure \ref{fig:4}(d3), and the corresponding tiling is shown in Figure \ref{fig:7}(1e).
We provide the proof of the Theorem  in the SI, section 1.
Whether or not the sufficient condition in Theorem \ref{thm:main} is also necessary is not clear. However, we believe that this condition can be relaxed:
\begin{conj}\label{con:main}
Let $M$ be a balanced, normal, convex tiling. Then there exists a combinatorially equivalent, soft polyhedric tiling $M'$.
\end{conj}

Theorem \ref{thm:main} shows, in broad terms, that soft tilings are ubiquitous in the combinatorial sense. Indeed, Figure \ref{fig:4}, panels (c3,c4,e1,e2,e3) show all five Dirichlet-Voronoi cells of point lattices  \cite{ghorvath_dirichlet, voronoi_dirichlet} with insets of the duals of the vertex polyhedra. In every case it is easy to build a Hamiltonian circuit.
We remark that, beyond the Dirichlet-Voronoi tessellations of point lattices, the condition set in Theorem \ref{thm:main} also proves to be sufficient for all uniform honeycombs. In fact, we are not aware of any monohedral tiling where the dual of the vertex polyhedron does not have either a Hamiltonian path (if the edge graph is simple) or a Hamiltonian circuit.

\subsection{Soft, softer, softest} 
\begin{figure*}[ht!]
\begin{center}
\includegraphics[width=\textwidth]{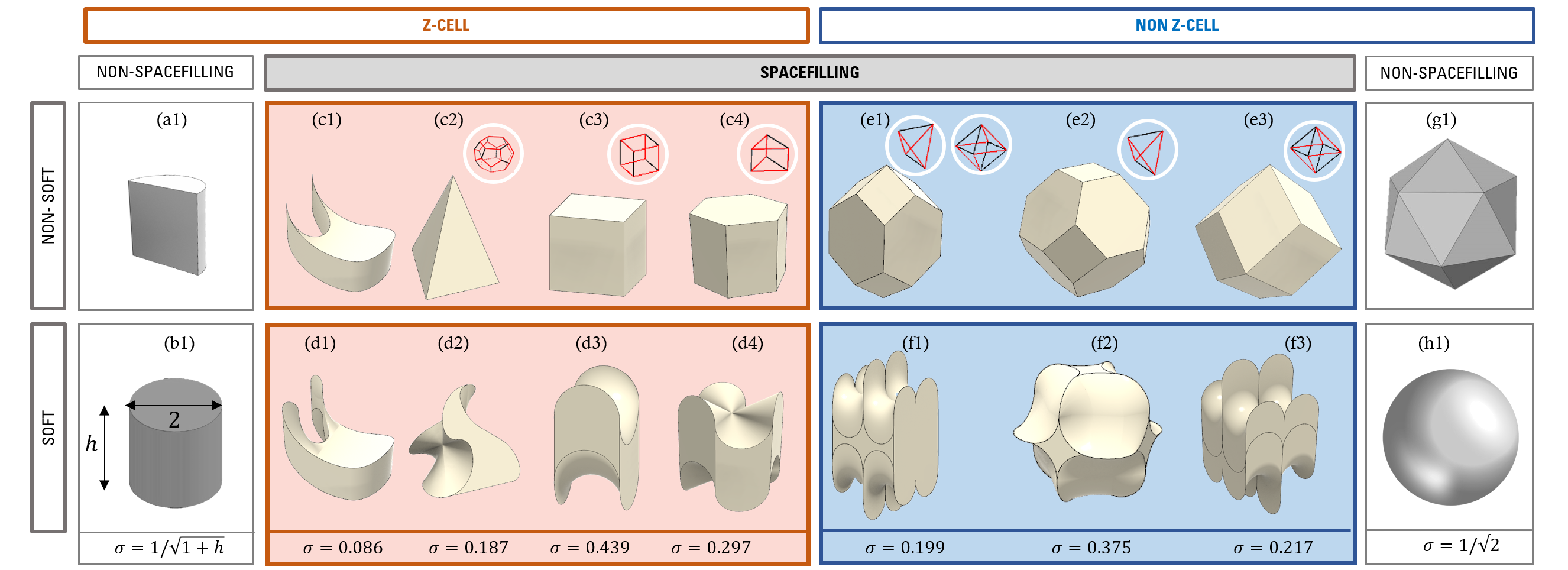}
\caption{Genesis of soft 3D cells. First row, panels a,c,e,g: non-soft cells. Second row, panels b,d,f,h: soft cells, with softness value indicated below each cell. Insets in white circles show the dual of the vertex polyhedron with Hamiltonian circuit indicated in red. II.  Individual panels: (a1) Half cylinder: non-soft, non-spacefilling cell. (b1) Cylinder. Maximal softness ($\sigma=1$) if $h=0$ (two-sided circular disc). (c1) Space-filling $z$-cell, resembling the Nautilus chamber. (c2-c4) Monohedral cells of prismatic tilings. (c2) Goldberg tetrahedron.  (c3-c4): Dirichlet-Voronoi lattice $z$-cells. (c3) cube (c4) hexagonal prism. (d1-d4): Soft versions of (c1-c4) as monohedric soft $z$-cells. (e1-e3) Dirichlet-Voronoi lattice  non-$z$ cells. (e1) Elongated dodecahedron. (e2): Truncated octahedron. (e3): Rombododecahedron.(f1-f3): Soft versions of (c1-c3) as monohedric, soft, non-$z$-cells. (g1) Non-spacefilling non-soft cell: Icosahedron. (h1) Non-spacefilling soft cell: sphere.}\label{fig:4}
\end{center}
\end{figure*}

The geometry of convex monohedral tilings is constrained: for example, the shape of the five Dirichlet-Voronoi cells of point lattices  \cite{ghorvath_dirichlet, voronoi_dirichlet}, illustrated in Figure \ref{fig:4}, panels (c3,c4,e1,e2,e3) is completely determined by the lattice. By using the edge bending algorithm introduced above, we constructed the corresponding soft monohedric tilings, the five soft cells of which are shown in Figure \ref{fig:4}(d3,d4,f1,f2,f3). (For more detail see SI, sections 2B and 4A.)  While the shape of the Dirichlet-Voronoi polyhedral cells is exactly determined by the lattice, the exact shape of the corresponding soft cells is not uniquely defined. 

To understand better the geometry of soft cells and to relate them to physical realisations, we introduce a  continuous scale on softness for 3D shapes. In the spirit of Blaschke's Rolling Ball Theorem \cite{blaschke}, we imagine that two identical circular disks are rolled along the opposite sides of a planar curve $c$ in such a manner that their contact point at $P_c \in c$ is always identical. We call the maximal radius of these disks under the condition that they can be both rolled smoothly locally at $P_c$ the \emph{rolling radius} of $c$ at $P_c$ and denote it by $\rho(P_c)$. If $c$ is at least $C^2$-smooth at $P_c$ then $\rho(P_c)$  is identical to the radius of curvature of $c$ at $P_c$ but if $c$ has a vertex at $P_c$, no finite disk can be smoothly rolled and we have $\rho(P_c)=0$.  We can generalize this concept to topological spheres $S$ embedded into 3D space as follows. Let $P_S$ be a point on the boundary of $S$. Then, the rolling radius $\rho(P_S)$ of $S$ at $P_S$ is defined as the supremum of the rolling radii of any planar section of $S$ at $P_S$. We call the infimum of rolling radii of all points $P_S$ on the boundary of $S$ the rolling radius $\rho(S)$ of $S$. We define the softness of $S$ as
\begin{equation}\label{eq:softness}
    \sigma(S)=\rho(S)\sqrt{\frac{2\pi}{A(S)}},
\end{equation}
where $A(S)$ is the surface area of $S$. If $S$ is at least $C^2$-smooth then $\rho(S)$ is the infimum of radii corresponding to the smaller principal curvature
and $\rho(S)=0$ as soon as $S$ has a sharp corner. As an example consider the family $S_a$ of oblate spheroids with semi-axes $(a,1,1)$ with $0 \leq a \leq 1$. Here we have $\rho(S_a)=1$ for every value of $a$ (see SI, Sec 3), so $\sigma(S_a)$ only depends on the surface area which is a monotonic function of $a$ as
    $\sigma(S_a)=\sqrt{e/(  e + a\, \text{artanh}(e))}$,
where $e=\sqrt{1-a^2}$ and $\text{artanh}(e)$ is the inverse hyperbolic tangent of the excentricity $e$.
The sphere has $\sigma(S_1)=1/\sqrt{2}$ and the circular disc has $\sigma(S_0)=1$ (see Figure \ref{fig:4} (b1)) and the latter case seams to be the global supremum of softness for all compact bodies. 

Using the above concept, we can now further deform a soft tiling  to maximize its softness. We found some rules (see Materials and Methods, subsection \ref{ss:softness}) which resulted in relatively high softness values. We compute and give the softness for the cells in Figure \ref{fig:4} (d1-d4,f1-f3) by using (\ref{eq:softness}).  It is not clear whether any of these cells have maximal softness but the soft monohedric cells in panels (d3,d4, f1,f2,f3) corresponding to the 5 Dirichlet-Voronoi cells on lattices appear to have softness close to the maximum if the lattice is fixed (see SI, section 4A for details on the computation). However, in case of (d3) and (d4), softness can be further increased by changing the lattice so that the distance between the two curved faces decreases.

\section{Soft cells in nature}
Establishing and maintaining sharp corners in physical cells is difficult and costly as surface tension and elasticity naturally tend to smooth corners. Hence, it is not surprising that many soft tilings are found in Nature. For instance,
soft $z$-cells appear to be ideal models for a variety of of natural shapes such as tip growth, one of the most ubiquitous biological shape evolution processes \cite{reinhardt1892wachsthum, goriely2008mathematical,goriely2017mathematics}. $Z$-cells also describe well the shape of blood cells travelling  through capilaries \cite{secomb1991red} (cf. panels (d2) and (d3) in Figure \ref{fig:3}). But, perhaps the most striking appearance of such structures is found in the  chambers of some seashells that we now explore.

\subsection{The soft geometry of chambered seashells}\label{ss:ammonites}
Chambered seashells are a fascinating feature found in certain mollusks, primarily in cephalopods. The most well-known examples include the extinct ammonites with shells divided into chambers connected by a tube, the \emph{siphuncle}, and the extant nautilus, well known for his iconic spiral divided into chambers. The animal lives in the outermost chamber and uses the siphuncle in order to control buoyancy by adjusting the gas-to-liquid ratio in the chambers. This is done by removing water from the chambers and replacing it with gas, which is mainly nitrogen, carbon dioxide, and argon in the case of the nautilus.

\begin{figure}[ht!]
\begin{center}
\includegraphics[width=\columnwidth]{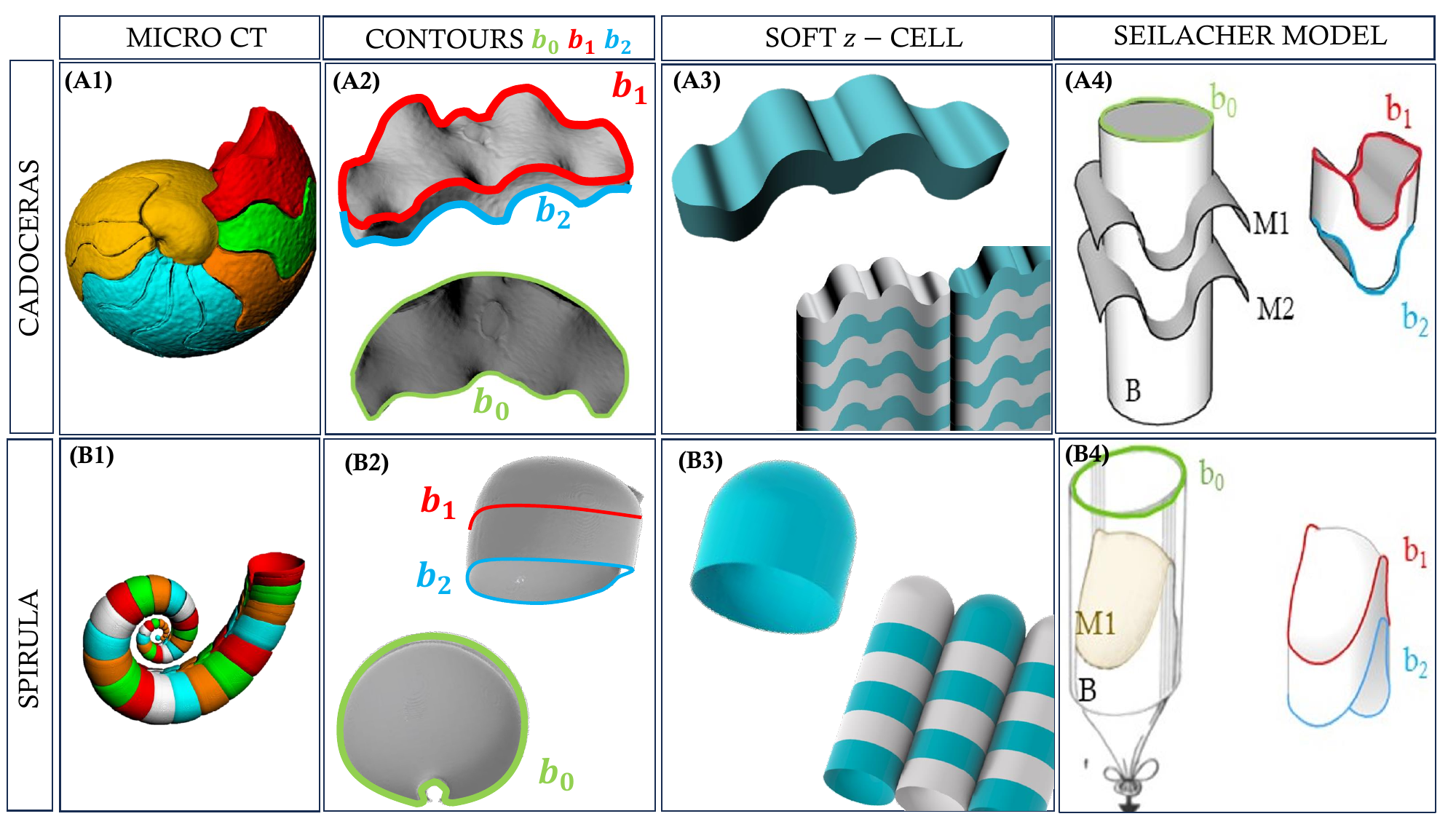}
\caption{Soft cells in chambered shells. Upper row: Ammonite shell chambers (Cadoceras) (A1) All chambers, reconstructed from micro CT dataset \cite{Lemanis}. (for details see Supplementary Information). (A2) Individual chamber with smooth upper and lower contours $b_1,b_2$ and base contour $b_0$. (A3) Soft, non-spacefilling $z$-cell as model of Cadoceras chamber.  (A4) Seilacher's paper model for septa: cylinder with base $b_0$ intersected by two smooth surfaces along curves $b_1,b_2$. \cite{Seilacher_1975}Bottom row: the chamber of the extant Spirula spirula shell. (B1) All chambers, reconstructed from micro CT dataset \cite{Lemanis}.
 (B2)   Individual chamber with smooth upper and lower contours $b_1,b_2$ and base contour $b_0$.  (B3) Soft, non-spacefilling $z$-cell as model of the Spirula chamber. (B4) Seilacher's balloon model: inflated balloons glued along the curves $b_1,b_2$ to the wall of cylinder with base $b_0$. \cite{Seilacher_1975}. }\label{fig:5}
\end{center}
\end{figure}

The chambers in segmented seashells grow under the constraint to fill the space provided by the outer shell. The geometry of the chambers has been investigated only recently, MicroCT technology providing unprecedented access to 3D images \cite{Lemanis, dundee} of which Figure \ref{fig:4} shows two examples.
The curved, soft contours of both chambers immediately catch the eye. The curved, spiral geometry of the tubular shell has no direct bearing on the softness of the chambers. Following Seilacher \cite{Seilacher_1973} we model the tubular shell as a prismatic object: the contour of the tube, orthogonal to the $z$ axis is the base $b_0$ of the prism in the geometric model (see also Figure \ref{fig:7}(1)).
Since the contour $b_0$  of shells tends to be smooth, if the tube is segmented in a generic fashion by a smooth interface  $M_1$ and its parallel $z$-translation $M_2$ (serving as the geometric model of septa), then the intersection lines $b_1,b_2$, defining a finite shell segment, are also smooth and the segments of the prism, modeling shell chambers emerge as soft $z$-cells (for a detailed discussion on the geometry of soft $z$-cells, see Materials and Methods, subsection \ref{ss:zcells}).  Geometric models of shell chambers were developed by Seilacher in a series of seminal papers \cite{Seilacher_1973, Seilacher_1975, Seilacher_1995}. Here we discuss two of his models, the \emph{paper model}  and the \emph{balloon model} describing the geometry of septal walls with vanishing and with positive Gaussian curvature, respectively. Despite the fact that these models are  qualitative, they both predict that the respective shell chambers will be soft $z$-cells. Each  model can approximate some aspects of the shell chamber's geometry, however, in some cases one of the models may offer a good approximation on its own and in Figure \ref{fig:5} we  examine such special shell chambers.

The first example is an  extinct ammonite genus (Cadoceras). Relying on Micro CT datasets presented in \cite{Lemanis} we traced the cross section $b_0$ and the spatial contours $b_1,b_2$, characterizing
the shape of the chamber. All three contours are smooth, and septal walls can be closely approximated by a surface with vanishing Gaussian curvature. Therefore they define a soft, non-space-filling $z$-cell (see Figure \ref{fig:5} (A1)-(A4)) the geometry of which is well approximated by Seilacher's paper model.

Our second example is the shell of an extant species, the deep water mollusk Spirula spirula. Here the septal walls resemble spherical caps with non-vanishing Gaussian curvature and they are well approximated by the balloon model. As before, the model of the chambers emerges as a soft, non-spacefilling $z$-cell (see Figure \ref{fig:5} (B1)-(B4)).

\subsection{The chamber of the Nautilus}\label{ss:nautilus}

\begin{figure*}[ht!]
\begin{center}
\includegraphics[width=\textwidth]{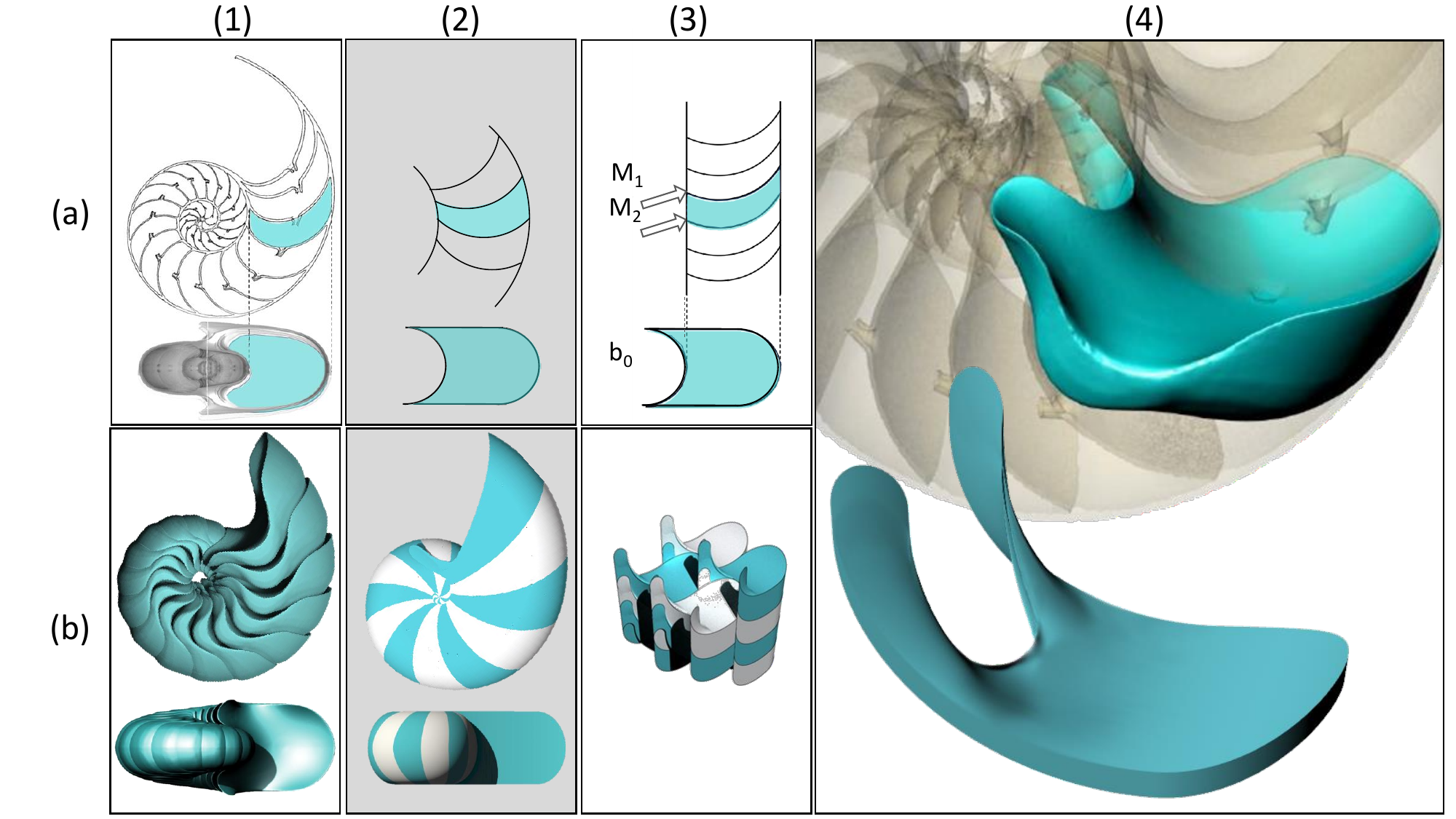}
\caption{The geometry of the Nautilus chamber. Row (a): Sections. Row (b): 3D geometry. Column (1)  Measured datasets, using data from \cite{dundee}. Column (2): Spiral model. Column (3): Prismatic model. Panel (4): Visual comparison between Nautilus chamber and its geometric model. For the geometric operations connecting (3a) to (3b) see Figure \ref{fig:7}, row (2).}\label{fig:6}
\end{center}
\end{figure*}

The most famous example of chambered seashells is undoubtedly the Nautilus which has been subject of geometric interest \cite{Nautilus_Hewitt, Nautilus_MOULTON} ever since the seminal book \cite{darcy} of D'Arcy Thompson appeared in 1917, pointing out that the \emph{2D geometry} of the section in the plane of symmetry  is a good approximation of the logarithmic spiral. Here we show that the \emph{3D  geometry} of the chamber is not only a good approximation of a soft $z$-cell (like other chambered shells) but it is also  a good approximation of a soft, space-filling, monohedric cell: Indeed, panel (4) of Figure \ref{fig:6} shows the striking visual resemblance between the Nautilus chamber \cite{dundee} and its soft, spacefilling geometric model.

The latter was created using two orthogonal sections of the shell, illustrated in Figure \ref{fig:6}(1a) and we simplified these two sections  in two steps (see Figure \ref{fig:6}(2a, 3a)) to obtain a geometric model.  In the first step  we assumed that the mostly flat shell is \emph{ideally flat}, i.e. it is constrained between two parallel walls. We call this the \emph{flatness assumption} and the resulting model the \emph{spiral model}. In the second step, following Seilacher \cite{Seilacher_1995}, we simplified the meridian section to be that of a straight prism and we call this the \emph{prismatic assumption} and the resulting model the \emph{prismatic model}.  The flatness assumption results in a soft, monohedric planar cell (cf. Figure \ref{fig:6}(2a) and Figure \ref{fig:3}(1d)) as the base $b_0$ of the prism in the prismatic assumption. We used the sections  shown in Figure \ref{fig:6}(3a) to create the 3D model model of the chamber  in two steps  (see Materials and Methods, subsection \ref{ss:zcells} and Figure \ref{fig:7}). In the first step the prism  is intersected by two parallel, cylindrical surfaces resulting in a softened, space-filling $z$-cell with two sharp corners. In the second step we use the edge bending algorithm to make this cell soft, resulting in a space-filling, soft cell shown next to the chamber in Figure \ref{fig:6}(4). Figure \ref{fig:6}(1b) shows how the chambers fill the shell, Figure \ref{fig:6}(3b) shows how the geometric model of the chamber fills 3D space and Figure \ref{fig:6}(2b) illustrates the geometry of the model without the prismatic assumption. As we can see, in the spiral model  we also have a soft, space-filling tiling, albeit not a normal tiling: due to the spiral geometry the tiling includes both infinitely small and infinitely large cells. We also remark that, unlike the examples in Figure \ref{fig:5}, the shape of the Nautilus chamber can only captured by combining two different Seilacher models (see SI, sec.5.).

\section{Summary and outlook}
We introduced a new class of shapes called \emph{soft cells} that tile 2D and 3D Euclidean space and have a minimal number of sharp corners. In 2D this implies having $v^{\star}=2$ corners while in 3D soft cells have no sharp corners. The lack of sharp corners  and their soft, highly curved geometry makes soft cells candidate models for biological structures which evolved under full or partial constraint to fill space. An important special class, representing partial constraint, are soft $z$-cells which fill  prismatic tubes without gaps. Similar constraints arise in biological growth and we can observe shapes strongly reminiscent of soft $z$-cells not only among blood cells travelling at high speed but also in tip growth and among chambers of seashells. In particular, the Nautilus shell not only exhibits chambers with the geometry of soft $z$-cells but these cells are remarkably close to space-filling $z$-cells.

Soft cells are not only abundant in Nature: we gave a sufficient condition under which a polyhedral tiling can be continuously transformed into a soft tiling. In fact, the proof (given in SI, section 1C and 1D) shows that a broader claim is also true: even if the original tiling is not polyhedral, if the vertex polyhedron's dual has a Hamilton circuit, the edge bending algorithm can be executed and the tiling can be turned into a soft tiling. This encompasses a broad class of tessellations, including, for example, tessellations by scutoids \cite{scutoid_gomez}. 

Beyond Nature and mathematics, we also find soft cells emerging in art. In  Figure \ref{fig:3} we showed examples
of 2D soft cellsappearing in the work of Zaha Hadid. However, there are other notable artists, like Katsushika Hokusai or Victor Vasarely,  who depictsoft 2D cells (see SI, section 4B). Even 3D soft cell seem to have roots in art, albeit in a curious manner. It not only appears that architects used such shapes to avoid corners, they even claimed to have identified the soft cell in panel (d3) of Figure \ref{fig:4} by a construction starting from the G\"omb\"oc shape \cite{archdaily}. In the SI, section 4B, we show this design along with another building which reflects the shape of the soft 3D cell in panel (d4) of Figure \ref{fig:4}.

\subsection{Open questions}
This new class of shapes suggests several questions, the answers to which could shed more light on their geometry and the connections to nature.  From the mathematical point of view, a proof of Conjecture 1 would be most desirable. However, there are other interesting questions related to the edge bending algorithm. While the latter always preserves the combinatorial structure of the tiling,  metric properties may or may not be preserved. Most notably, the number of congruence classes for the tiles may change and if the initial tiling was monohedral, it may not be possible to produce a monohedric soft tiling by this algorithm. Since we were particularly interested in soft monohedric tilings, the examples in this paper
illustrated cases where the algorithm preserved the monohedric property of the tiling. This is always possible, as long as we have a lattice-like tiling with a single cell as fundamental domain. This condition is certainly not necessary, the tetrahedral soft cell in Figure \ref{fig:4} (d2) is part of a monohedric tiling which is not lattice-like (full tiling shown in SI, Section 4, Figure S17). Any necessary condition appears to be nontrivial, since there are many monohedral tilings, the soft version of which is not monohedric. The simplest such example is probably a planar triangulation,
created by a rectangular grid with all diagonals added.
Other examples can be found among monohedral pentagon tilings  \cite{pentagons} and one most notable example is the recently discovered aperiodic monotile ('einstein') \cite{einstein_hat} and its chiral version \cite{einstein_spectre}. Both have 13 sharp corners the number of which can not be reduced if we want to preserve the monohedric property of the tiling. Hence we pose the question what is minimal number of sharp corners for an aperiodic monotile.

From the point of view of applications, we showed examples where 2D soft cells helped to describe natural patterns and we found several natural phenomena where 3 dimensional soft $z$-cells appear to describe the essence of the geometry.  Still, we not yet found any example of a soft, 3D non-$z$-cell appearing in Nature, although,
as Figure \ref{fig:4} illustrates, these attractive shapes appear to be abundant in geometry. We believe that finding an example of a soft non-$z$ cell in Nature would greatly help to understand the connection between growth processes and this new class of shapes.


\section{Materials and Methods}


\subsection{Construction of $z$-cells}\label{ss:zcells}
\begin{figure}[ht!]
\begin{center}
\includegraphics[width=\columnwidth]{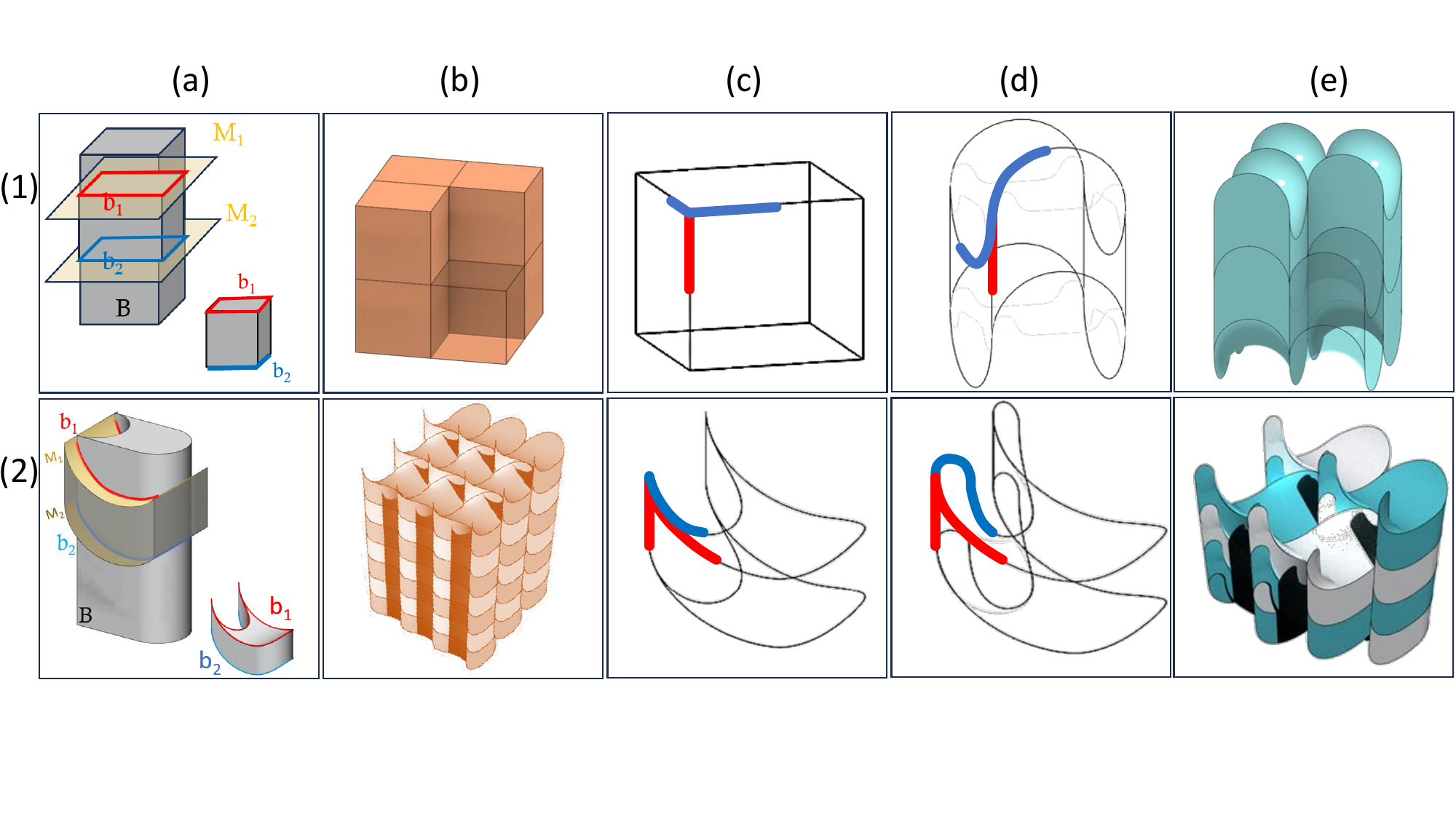}
\caption{Geometry of $z$-cells and the edge bending algorithm. Row (1): Cubic cell. Row (2): Nautilus cell. Columns:
(a) Genesis of $z$-cell: intersecting of prism $B$ with manifolds $M_1,M_2$ produces $z$-cell with respective lower and upper contours $b_1,b_2$.(b): prismatic tiling by $z$-cells. (c) Wireframe model of $z$-cell. Red line marks specified direction at node, blue lines mark edges not aligned with the red direction. (d) Spcefilling $z$-cell transformed into soft, spacefilling $z$-cell by edge bending. Red edges did not change, blue edges have been bent. (e): prismatic tiling by soft $z$-cells.}\label{fig:7}
\end{center}
\end{figure}

As stated in the main text, $z$-cells are constructed by the segmentation of infinite prisms with axis $z$ into identical, finite parts. Here, we only consider the case where adjacent $z$-cells are related by $z$-translation. (For the general case see SI, Sec.3A.)  Figure \ref{fig:7} shows two such examples: the cubic cell (upper row) and the geometric Nautilus cell (bottom row), both also shown in Figure \ref{fig:4}, panels (c1-d1) and (c3-d3), respectively. We consider a smooth manifold $M_1$ which intersects the prism $B$ with base $b_0$ in such a manner that it has exactly one transverse intersection point with every straight line on the boundary of $B$ and we call the curve consisting of these points $b_1$. The manifold $M_2$ is a $z$-translation of $M_1$ and intersects $B$ in the curve $b_2$ which is the $z$-translation of $b_1$. The curves $b_1,b_2$ define a finite segment $\mathcal{B}$ of the prism $B$ and we call this segment a $z$-cell. If both the base $b_0$ of the prism and the manifold $M_1$ are smooth then $b_1,b_2$ are also smooth and $\mathcal{B}$ will be a soft $z$-cell.

\subsection{Edge bending}\label{ss:edgebending}
In the edge bending algorithm we consider a non-soft space-filling cell with sharp corners as initial condition.
At each sharp corner, we specify one straight line and bend all edges in such a manner that, on one hand, their half-tangents become aligned with the specified line, on the other hand, there is at least one half tangent associated with both directions defined by the line. Most conveniently, the specified line coincides with one of the half-tangents. While several edges become curved in this process, the combinatorial structure of the cell is not changed. To maintain the space-filling property of the cell we have to pick the same straight line for every sharp vertex meeting at any node. In the case of the cubic cell (Figure \ref{fig:7}, upper row) it implies that we have to pick parallel lines at every vertex of the cell since all vertices meet at one node.
\subsection{Achieving high softness values}\label{ss:softness}
 To achieve high softness in the edge bending process, at sharp vertices connected by an edge we always specify two co-planar lines $e$ and $f$. Subsequently, we replace the edge by a Dubins path \cite{dubins} under the constraint that the fixed curvature should be minimal. The lines $e$ and $f$  define four infinite planar domains in the generic case and three domains if they are parallel and the original straight edge is contained in exactly one of those domains. We construct the Dubins path under the additional constraint that it should also be contained in the same planar domain as the original edge. After tracing the edges by this algorithm, we select the set of (possibly curved) edges which correspond to the boundary of a face and  construct a minimal surface on each face. The soft cells in panels (d2-d4), (f1-f3) have been constructed by this algorithm. Whether or not they have maximal softness is not clear, with the possible exception of (d3) and (d4).

\section{Acknowledgement}
The authors thank Lajos Czegl\'edi  for his invaluable help with 3D data rendering and presentation.
\textbf{Funding:}
KR and GD: This research was supported by NKFIH grants K134199 and EMMI FIKP grant VIZ. KR: This research has been supported by the program UNKP-23-3 funded by ITM and NKFI. The research was also supported by the Doctoral Excellence Fellowship Programme (DCEP) funded by ITM and NKFI and the Budapest University of Technology and Economics. The gift representing the Albrecht Science Fellowship is gratefully appreciated. AG:
This work was made possible through an International Exchanges Scheme Award from the Royal Society.

\bibliographystyle{unsrt}
\bibliography{soft}

\end{document}